\documentclass[prl,twocolumn,showpacs, amsmath,amssymb,aps]{revtex4-1}
\usepackage{graphicx}
\usepackage{bm}

\begin{document}
\newcommand{\balpha}{\mbox{\boldmath$\alpha$}}

\title{Surface plasmon amplification by stimulated emission of radiation in hyperbolic metamaterials}

\author{Vitaliy N. Pustovit, Augustine M. Urbas and David E. Zelmon}

\affiliation{Materials and Manufacturing Directorate, Air Force Research Laboratory, Wright Patterson Air Force Base, Ohio 45433, USA}


\begin{abstract}
 We study theoretically and numerically spasing conditions and optical dynamics of a composite hyperbolic metamaterial interacting with gain molecules. By combining Maxwell-Bloch equations with Green's function approach, we calculate lasing frequency and threshold population inversion for various gain density in the gain layer. We demonstrate high level of enhancement for photonic density of states in spacious spectral range provided by hyperbolic metastructures.  We find that direct dipole-dipole interactions between molecules in the gain layer has a negligible effect on spasing conditions. We identify a region of parameters in which spasing can occur considering these effects.
 \end{abstract}


\maketitle
\section{Introduction}
Enhanced spontaneous emission of nuclear magnetic moments in the presence of a resonant circuit was predicted by Purcell in 1946 \cite{Purcell}.  Purcell's calculations suggest that the properties of any emitter may be modified by its material environment. Over the last few decades there have been many attempts to control the spontaneous emission rate of organic dyes and quantum dots by positioning them near the surface of materials with a high optical density of states (DOS) \cite{Dulkeith,Lakowicz,Dulkeith2,Novotny, Kuhn,Seelig}. Microcavities and photonic crystals are among the most promising systems currently available for enhancing spontaneous emission. However, these approaches are significantly restricted in their practical use.  The very sharp resonances required to see the Purcell effect set restrictions on the spectral width of potential emitters and compatibility with different sources. Plasmonic systems based on metal nanoparticles are problematic because of the difficulty controlling size, shape and surface chemistry of the particles during the formation of complicated nanostructures.  Artificial metamaterials with hyperbolic dispersion may overcome these limitations by supporting a large number of electromagnetic states that can couple to an array of quantum emitters leading to a broadband Purcell effect \cite{Narimanov, Liu, gu}. The Purcell effect in hyperbolic systems gives rise to nonradiative hyperbolic modes with large losses in microcavities. When combined with active gain materials such as quantum dots or dye molecules, conditions are created which concentrate light into mode volumes that are no longer limited by diffraction, allowing the formation of lasers or spasers (surface plasmon amplification by stimulated emission of radiation) \cite{Stockman,Stockman2,Bergman, josab, pustovit, noginov}. Such devices have already been explored in a limited number of materials and geometric configurations.
By combining the Maxwell-Bloch equations with Green function formalism we have derived spasing conditions for simple layered structures of a gain media composed of a doped dielectric and an HMM composed of a multilayer structure. The analysis divides the system behavior into regions where lasing, spasing and normal emission occur.

\section{Hyperbolic materials with high photonic density of states}
The ability to design nanostructured metamaterials has led to the development of many types of devices with unique optical properties which arise from intrinsic material properties, geometric configuration of the materials and near field coupling between properly designed subwavelength building blocks. Among the applications already imagined for metamaterials are subwavelength resolution imaging, cloaking devices and perfect absorbers \cite{Liu, jacob}. A class of metamaterials which has recently emerged as one of the most important at optical and near-optical frequencies are hyperbolic metamaterials (HMM). These structures can be made in such a way as to have dielectric tensors which are identical in form to those of ordinary uniaxial materials with one important difference: the dielectric tensor elements have opposite signs i.e. $\epsilon_{x}\epsilon_{z}<0$.  The most important property of such media is their intrinsic ability to support propagation of waves with large magnitude wave vectors $k$. In a vacuum, such waves are evanescent and decay exponentially. If we consider TM polarized modes in natural materials, the isofrequency relation is
 \begin{equation}
\frac{k_x^2+k_y^2}{\epsilon_x} + \frac{k_z^2}{\epsilon_z}= \frac{\omega^2}{c^2},
\end{equation}
If $\epsilon_x$ and $\epsilon_z$ have the same sign, this is the equation of an ellipsoid and the density of states is proportional to the volume between the ellipsoids at $\omega$ and $\omega + \Delta\omega$, which is finite. However, if $\epsilon_{x}\epsilon_{z}<0$ means that the isofrequency surface opens into open hyperboloid, hence the name hyperbolic metamaterials. The enclosed volume between two isofrequency surfaces is a measure of the photonic density of states of the system.  Since this volume diverges for HMMs so that, in the ideal limit, these materials can support an infinite photonic density of states. In real devices, the photonic density of states is extremely large but nonetheless finite due to nonvanishing losses in the material.  Fermi's golden rule states that the spontaneous emission lifetime of emitters such as fluorescent molecules or quantum dots is strongly affected by the photonic density of states. If the emitter is located in close proximity to a HMM, the result can be a decrease of state lifetime (absorption enhancement or quenching) or, if the device is operated in the saturation regime where the output signal is independent of the input power, photoluminescent enhancement. The measure of radiative enhancement is known as the Purcell factor defined as $F_p=\frac{\Gamma_r}{\Gamma^0_r}$ where $\Gamma_r$ is an enhanced radiative decay rate of emitter near a metamaterial and $\Gamma^0_r$ is a decay rate of the emitter in vacuum.
Divergence of the photonic density of states does not necessarily mean that large fluorescent enhancement would be observed since many of these are modes which do not couple to the vacuum \cite{couple1, couple2, couple3}. The problem of emission uncoupling for hyperbolic devices is critical and requires multiple steps to explore by practical fabrication different types of diffraction gratings aimed to uncouple enhanced emission \cite{strangi,menon}.

\section{Optical dynamics of composite HMM}
\begin{figure}
 \centering
 \includegraphics[width=0.8\columnwidth]{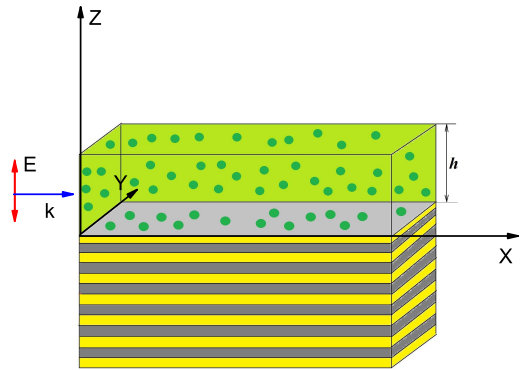}
 \caption{\label{fig:hmm} (color online) Multilayer HMM (silver/silica) with dye mixed PMMA layer on top}
  \end{figure}

We study the optical dynamics of a composite HMM system representing a silver/silica multilayers and layer of randomly distributed dyes (fluorescent molecules or quantum  dots) added on top of the HMM structure and imbedded in PMMA (polymethyl methacrylate) layer of thickness $h$ [see Fig.\ \ref{fig:hmm}].
 The system is illuminated by an incident TM polarized wave, assuming propagation of plasmonic modes inside HMM along the $x$ axis. In the standard  density matrix approach, the active molecules can be, to a first approximation, described by a pumped two level system  located at $\textbf{r}_{j}$ with excitation frequency $\omega_{12}$ between levels 1 and 2. Each molecule  is characterized by  polarization $\rho_{j}\equiv \rho_{12}^{(j)}$, molecule dipole moment $\textbf{p}_{j}=\mu\textbf{e}_{j}\rho_{j}$ and population inversion $n_{j}\equiv\rho_{22}^{(j)}-\rho_{11}^{(j)}$, where $\rho_{ab}^{(j)}$ ($a,b=1,2$) is  density matrix for $j$th molecule. For the mode of steady state operation and in the rotating wave approximation, the internal molecule dynamics is described by  optical Bloch equations \cite{haken, chipul}

%
\begin{align}
\label{maxwell-bloch}
& \left[i + \tau_{2}(\omega - \omega_{21}) \right] \rho_{j} = \frac{\tau_{2}}{\hbar} A_{j}\, n_{j}
\\
&n_{j}-n_0 = \frac{2i\tau_1}{\hbar}\left( A_{j}\rho_{j}^{*}- A_{j}^{*} \rho_{j}\right),
\nonumber\\
&n_0=\frac{W \tilde\tau_1 -1}{W \tilde \tau_1 +1},~~~~\tau_1=\frac{\tilde \tau_1}{W \tilde \tau_1 +1},
\nonumber
\end{align}
Here $\tau_2$ and  $\tilde{\tau_1}$ are the time constants describing phase and energy relaxation processes due to the interaction with a thermostat,  $W$ is the pump rate, and $A_{j}=\mu\textbf{e}_{j}\cdot \textbf{E}(\textbf{r}_{j})$ is an interaction term, defining the slow varying amplitude of the local field at the point of $j$th molecule ($\mu$ and $\textbf{e}_{j}$ are the molecule dipole matrix element and its orientation). The field  $\textbf{E}(\textbf{r}_{j})$ is created by all dipoles in the presence of the HMM, and it satisfies the following wave equation
\begin{align}
\label{maxwell}
{\bm \nabla} \times {\bm \nabla} \times {\bf E}({\bf r})
&-\epsilon ({\bf r},\omega )\frac{\omega^{2}}{c^2} {\bf E}({\bf r})=\frac{4\pi \omega^{2}}{c^2}\sum_{j}\textbf{p}_{j}\delta(\textbf{r}-\textbf{r}_{j}),
\end{align}
where $\epsilon({\bf r},\omega )$ is local dielectric function in the region of interest, $c$ is speed of light. The solution of Eq.~(\ref{maxwell}) can be written as \cite{pustovit}
\begin{eqnarray}
\label{electric}
 \textbf{E}({\bf r} ) = \textbf{E}_{0}({\bf r} ) + \frac{4\pi \omega^{2}\mu}{c^2} \sum_{j=1}^{M} \textbf{G}({\bf r},{\bf r}_{j}) \cdot \textbf{e}_{j}\,\rho_{j} ,
\end{eqnarray}
where ${\bf E}_{0}({\bf r})$ is the solution of homogeneous Eq.~(\ref{maxwell}) (i.e., in the absence of emitters but in the presence of HMM) and  $\textbf{G}({\bf r},{\bf r}')$  is its dyadic Green function.

The variety of the Green function techniques \cite{jackson,carbotte} have the advantage that for a particular given geometry, the solution has to be done only once; the problem involves an integral over the source and can be used to calculate the generated fields from any source distribution. We will construct a complete description of the optical response of the system based on the Green's function method characterizing the surrounding media. This will allow us to investigate the high gain coverage of the hyperbolic nanostructure with either an arbitrary distribution of emitters taking into account not only interactions between the hyperbolic system and the emitters but also between the emitters themselves. Although this formulation holds for arbitrary boundary conditions, exact analytical evaluation of the Green function tensor can be very cumbersome \cite{martin, martin2}. However, for certain geometries, the expression for the Green tensor can be significantly simplified.  The mathematical approach to this problem for the case when emitter's dipole oriented perpendicular to the surface of hyperbolic is shown in Appendix.
Finally, using Eq. (\ref{electric}) to eliminate the electric field, the system (\ref{maxwell-bloch}) takes the form
\begin{align}
\label{maxwell-bloch-hom}
&\sum_{k=1}^{M}\left [\left (\omega-\omega_{21}+i/\tau_{2}\right )\delta_{jk}- n_{j}D_{jk}\right ]\rho_{k}=0,
\nonumber\\
&n_{j}-n_0 + 4\tau_1\text{Im}\sum_{k=1}^{M}\left[ \rho_{j}^{*}D_{jk}\rho_{k}\right]=0.
\end{align}
where $\delta_{jk}$ is Kronecker symbol and $D_{jk}(\omega)$ is a frequency-dependent matrix in position space given  by
\begin{equation}
\label{matrix}
D_{jk}=\dfrac{4\pi\omega^{2}\mu^{2}}{c^{2}\hbar}\,\textbf{e}_{j}\cdot  \textbf{G}({\bf r}_{j},{\bf r}_{k}) \cdot \textbf{e}_{k}.
\end{equation}
Taking solution of the first equation for density of states $\rho_{j}$ and using it in the second equation, we can get:
\begin{equation}
\label{maxwell-bloch-hom2}
n_{j}\left [ 1 + 4 \tau_1 \tau_2 L(\omega) \sum_{k=1}^M |D_{jk} \rho_{k}|^2 \right]=n_0,
\end{equation}
where we introduce normalized lineshape function $L(\omega)$, which is in this case a Lorentzian,
\begin{equation}
\label{lorentzian}
L(\omega)=\frac{(1/\tau_2)^2}{(\omega-\omega_{21})^2 + (1/\tau_2)^2}
\end{equation}
The system eigenstates are determined by the spatial matrix $D_{jk}$ that incorporates the effects of molecular coupling to the hyperbolic structure. We now introduce the eigenstates of $D_{jk}$ as $D|J\rangle=\Lambda_{J}|J\rangle$, where $\Lambda_{J}=\Lambda'_{J}+i\Lambda''_{J}$ are the corresponding complex eigenvalues, and define a new collective variable as
\begin{equation}
\rho_{J}=\sum_{j}\langle \bar{J}|j\rangle \rho_{j},
~~
n_{JJ'}=\sum_{j}\langle \bar{J}|j\rangle n_{j}\langle j|J'\rangle,
\end{equation}
where, to ensure orthonormality of the basis, we introduced eigenstates $|\bar{J}\rangle$ of complex conjugate matrix $\bar{D}_{jk}$.
Finally, multiplying first equation in system (\ref{maxwell-bloch-hom}) and Eq.(\ref{maxwell-bloch-hom2})  both sides by $\langle \bar{J}|j\rangle$ and then summing over $j$, we get:
\begin{align}
\label{maxwell-bloch-hom3}
\sum_{J'=1}^M \left [(\omega-\omega_{21} + i/\tau_2)\delta_{JJ'} - n_{JJ'} \Lambda_{J'} \right]\rho_{J'}=0 ,
\nonumber\\
N=\frac{N_0}{1 + 4\tau_1\tau_2 L(\omega) \sum_{J=1}^M |\Lambda_J \rho_J|^2},
\end{align}
where $N=\sum_j n_j$ is the ensemble population inversion and $N_0=n_0 M$. Then, by connecting trace of density matrix with intensity
of the incident light as $I=\sum_J |\rho_J|^2$ and introducing saturated intensity $I_s$ as:
\begin{align}
\label{saturation}
I_s=\left [\frac{4\tau_1 \tau_2 \sum_{J=1}^M |\Lambda_J \rho_J|^2}{\sum_{J=1}^M |\rho_J|^2}\right]^{-1},
\end{align}
we obtain conventional expression for the stationary inversion
\begin{align}
\label{saturation2}
 N=\frac{N_0}{1+ \frac{I}{I_s} L(\omega)},
\end{align}
We also assume sufficiently large ensemble of dyes with relatively weak inhomogeneity that allows to adopt $n_{JJ'}=n \delta_{JJ'}$ and where
$n=N/M$ is the average population inversion per molecule. From the first equation of system (\ref{maxwell-bloch-hom3}) we can derive the characteristic equation for each mode,
\begin{align}
\label{maxwell-bloch-hom-coll1}
\omega-\omega_{21}+i/\tau_{2}-n\Lambda_{J}(\omega)=0,
\end{align}
indicating that each eigenmode acquires self-energy $\Lambda_{J}(\omega)$ due to interactions with HMM and each other. The resonance frequency of mode $J$ can be found from the real part of Eq. (\ref{maxwell-bloch-hom3}),
\begin{eqnarray}
\label{frequency}
\omega= \omega_{21} + \frac{1}{\tau_2}\frac{\Lambda'_{J}(\omega)}{\Lambda''_{J}(\omega) }.
\end{eqnarray}
while its imaginary part would define loss compensation condition,
\begin{eqnarray}
\label{loss_condition}
 n\tau_{2}\Lambda''_{J} =1,
\end{eqnarray}
and also determines the average population inversion \textit{per molecule}, $n$. Note, that Eq.(\ref{frequency}) and Eq.(\ref{loss_condition}) are valid for any plasmonic system with weak inhomogeneity of gain population inversion and in general determine specific spasing conditions in such systems.

\section{Numerical Results and Discussion}
Below we present the results of numerical calculations for an ensemble of dyes randomly distributed on top of HMM. We assume that all dyes are distributed randomly over the surface of HMM and, for simplicity, their dipole moments are oriented normally to the surface in accordance with incident TM field.
 \begin{figure}
 \centering
 \includegraphics[width=0.8\columnwidth]{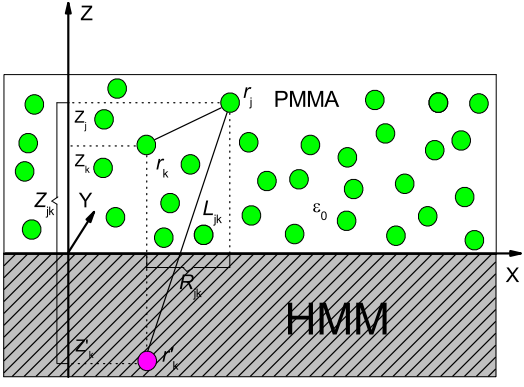}
 \caption{\label{fig:geometry} (color online) Geometry of HMM structure with dye mixed PMMA located on top.}
  \end{figure}
In this case, the matrix elements of $D_{jk}=D_{jk}^{0}+D_{jk}^{nr}$
\begin{eqnarray}
\label{green}
 D_{jk}^{0}(\omega)=-
 \frac{\mu^{2}}{\hbar}
\frac{(1-\delta_{ij})
\varphi_{jk}}{ r^3_{jk}},
\nonumber\\
 D_{jk}^{nr}(\omega)= \frac{\mu^{2}S}{L_{jk}^3 \hbar} \left[3\frac{Z_{jk}^2}{L_{jk}^2} -1\right]
 \end{eqnarray}
where $ D_{jk}^{0}$ represents the Coulomb interaction \cite{pustovit} and $D_{jk}^{nr}$ the non-radiative interactions, $\gamma_{jk}$ is the angle between dye locations ${\bf r}_j$ and ${\bf r}_k$, and $\varphi_{jk}=1+ \sin^2 (\gamma_{jk}/2)$.
 In second term, $Z_{jk}=z_j+z_k$ are z-coordinates of dipoles placed on the surface and $R_{jk}=|\rho_j-\rho_k|$ is a lateral distance between two dipoles in $XY$ plane with $\rho_j=\sqrt{x_j^2 +y_j^2}$. We have introduced the characteristic length $L_{jk}=\sqrt{Z_{jk}^2+R_{jk}^2}$ as depicted on Fig.\ \ref{fig:geometry} and parameter $S$ which is the normalized effective polarizability of an image dipole in the upper half-space (see Appendix)
\begin{eqnarray}
S= \frac{\epsilon_x-\epsilon_0 \sqrt{\frac{\epsilon_x}{\epsilon_z}}}{\epsilon_x+\epsilon_0 \sqrt{\frac{\epsilon_x}{\epsilon_z}}},
\end{eqnarray}
We assume that the HMM is uniaxial with the dielectric tensor containing two independent non-zero elements which are determined using the Maxwell-Garnett formulas \cite{sipe, sipe2}
\begin{eqnarray}
\epsilon_x=\epsilon_y=f \epsilon_m + (1-f)\epsilon_d,
\nonumber\\
\epsilon_z=\left[f\epsilon^{-1}_m + (1-f) \epsilon^{-1}_d\right]^{-1},
\end{eqnarray}
with $f=d_m/(d_m+d_d)=0.5$ set as a filling fraction of metal. Here $d_m$ and $d_d$ are average layer thicknesses of metal and dielectric respectively, $\epsilon_0 = 2. 2$ is the dielectric constant of a dye mixed in PMMA. Generally, effective medium approximation (EMA) may  overestimates the Purcell effect unless the thickness of the metallic layers is rather small \cite{sipe, sipe2}, but it is widely used in the HMM literature as a first step  approximation and may serve for the purposes of this paper where HMM is a source of high density of states.

\begin{figure}
 \centering
 \includegraphics[width=\columnwidth]{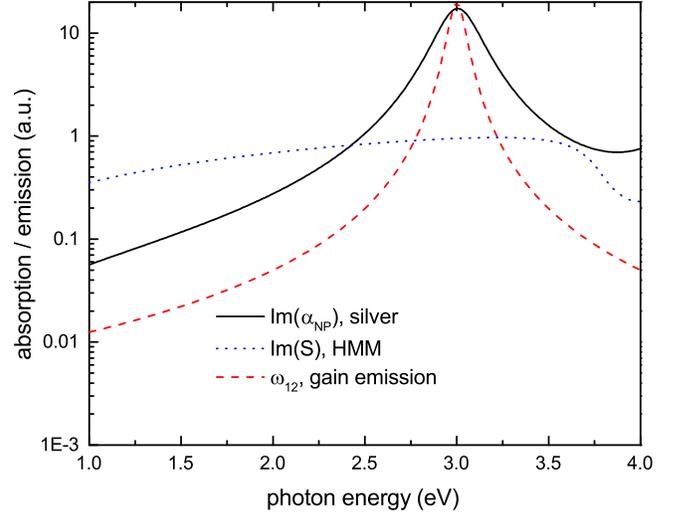}
 \caption{\label{fig:alfa} (color online) Comparison of effective polarizability of an image dipole in upper half-space above HMM with normalized on radius polarizability for silver nanoparticle and with dye's emission spectra centered at silver surface plasmon $\omega_{12}=3$ eV}
  \end{figure}

The eigenstates were found by numerical diagonalization of matrix $D_{jk}$ in configuration space and the spasing state was determined as the one whose eigenvalue $\Lambda_{s}(\omega)$ has the largest imaginary part $\Lambda_{s}''(\omega)$ in order to satisfy the spasing threshold condition (\ref{loss_condition}). In contrast with the spaser formed with a spherical nanoparticle \cite{pustovit}, the planar structure of the HMM removes any degeneracies contained in $|J\rangle$ eigenstates, reducing the generally multipolar mode interactions to the simple dipole ($l=1$) coupling between the emitters and the HMM.
The nature of plasmon coupling between gain and multilayered metal/dielectric metamaterial requires some special attention. It has been shown \cite{podolskiy, nam, nam2} that volumetric (or bulk) plasmons that reside at the internal interfaces of the HMM are not a well known SPP's that appears at the simple interface between metallic film and air, for example. The plasmon modes propagated inside lamellar layers would have different excitation conditions and maintain the largest contribution into the local density of states. This prompts us to conclude that stimulated emission of spaser in HMM \cite{noginov2}  is, indeed, enhanced by volumetric plasmons. Comparing with conventional spaser  \cite{noginov, Stockman, Stockman2}, we find out that area of plasmonic enhancement (density of states) specific for the HMM is much wider than localized nanoparticle spaser. That fact provides opportunity to resolve spasing for much larger wavelength range (see Fig.\ \ref{fig:alfa}).

\begin{figure}
 \centering
 \includegraphics[width=1\columnwidth]{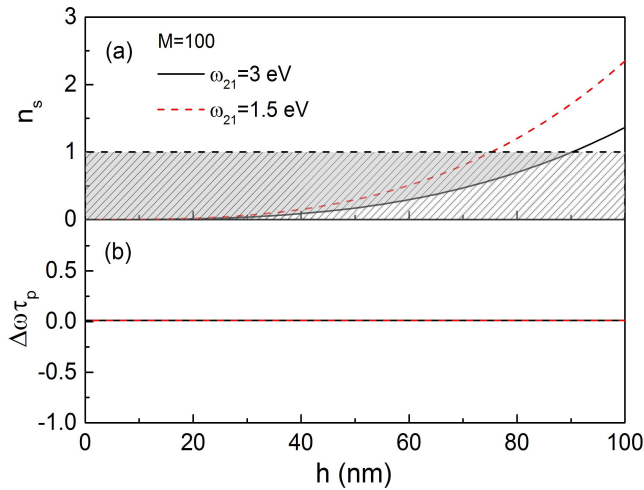}
 \caption{\label{fig:M100} (color online) (a) Spasing threshold $n_s$, where the hatched region represents the gain condition and the spasing region is shaded gray, and (b) frequency shift $\Delta \omega=\omega_s-\omega_{21}$ for M=100 molecules are plotted vs thickness of the PMMA gain layer.}
  \end{figure}

\begin{figure}
 \centering
 \includegraphics[width=1\columnwidth]{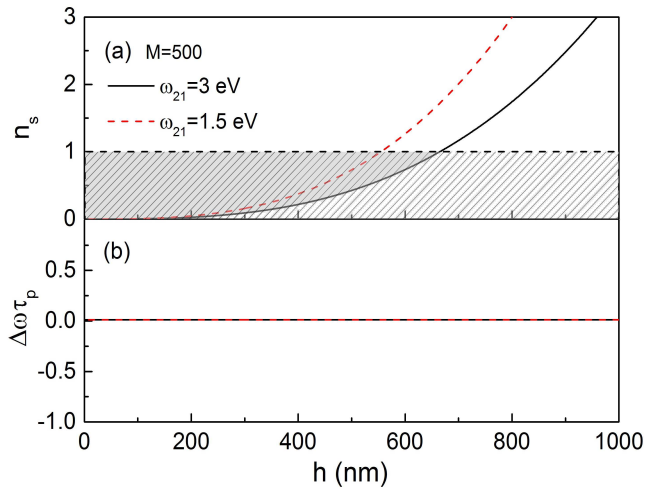}
 \caption{\label{fig:M500} (color online) (a) Spasing threshold $n_s$, where the hatched region represents the gain condition and the spasing region is shaded gray, and (b) frequency shift $\Delta \omega=\omega_s-\omega_{21}$ for M=500 molecules are plotted vs thickness of the PMMA gain layer.}
  \end{figure}

In our simulations, we consider two cases in which either $M = 100$ (case 1) or $M=500$ (case 2) Rhodamine 6G molecules are randomly distributed over the surface ($100\times 100$ nm) of a HMM which consists of a stack of alternating silver/silica layers with all layers having the same thickness and the dye dipoles directed perpendicular to the plane of the stack. The bandwidth, dipole matrix element and dye relaxation time were taken as $\hbar/\tau_2=0.06$ eV, $\mu=4$ D and $\tau_1 \sim 10^{-8}$ seconds respectively, all of which are typical values for the Rhodamine family of dyes. The material parameters for the Drude model for bulk silver dielectric function were taken from \cite{christy}. Using the slowly varying approximation in the vicinity of resonance of HMM the frequency of spaser $\omega_{sp}$ is obtained from solution of  Eq. (\ref{frequency}), which results in two conditions for spasing. First, the spasing frequency should lies well within the plasmon spectral band, i.e. $-1\leq \bigtriangleup \omega \tau_p \leq 1$ \cite{pustovit}. The second restriction comes from the traditional lasing requirement on compensation of all losses given by Eq. (\ref{loss_condition}) and defines a "lasing threshold" which requires that average ensemble population inversion per molecule not
exceeds the unity, i.e. $n_{s}\leq 1/{\tau_2 \Lambda''_J}$. Here, among $J=1..M$ modes we selected only one super-fast lasing "supermode"  which has maximal value of imaginary part and, as concluded in \cite{josab}, corresponds to the lasing state.  In Fig.\ \ref{fig:M100}  we present calculations for $M = 100$ dyes with dipoles randomly distributed on the top of HMM and oriented perpendicular to the surface  [see Fig.\ \ref{fig:M100}]. We show resonance frequency shift $\Delta \omega=\omega_s-\omega_{21}$, normalized by plasmon lifetime $\tau_p$, and threshold population inversion per molecule $n_s=N_s/M$ as a function of gain layer thickness $h$ for $M=100$ gain molecules.
The two types of calculations have been performed at two different gain emission frequencies which reveals that the calculated $\Delta \omega$ is nearly vanishing, while $n_s$ increases with $h$ before reaching its maximum values allowed for spasing at $h=75$ nm and $h=90$ nm for $\omega_{21}=1.5$ eV and $\omega_{21}=3$ eV, correspondingly. One may also notice here a high sensitivity of spasing threshold to the position of the dye's emission frequency in respect of the silver plasmon resonance. Interestingly, that contribution from the Coulomb interaction (see first term in Eq.\ref{green}) appeared to be quite small comparing with nonradiative term and negligible to make any influence in observed results. This fact is even more important as it was found in \cite{pustovit}, in case of a classical spaser based on metal nanoparticles, the random Coulomb shifts may lead to significant detuning and loss of coherency in the spaser emission. In Fig.\ \ref{fig:M500} we repeat our calculations for a larger number of gain molecules, $M=500$, which demonstrate the high level of sensitivity of the lasing threshold to the amount of applied gain. The required concentration of gain needed for spasing can be achieved even for a dye layer of approximately $600$ nm thickness when $M=500$ molecules randomly distributed inside.

In conclusion, we performed a theoretical and numerical study spasing of conditions of hyperbolic metamaterials interacting with gain media. With the help of a two level Maxwell-Bloch model and the Green's function approach, we estimated the spasing threshold for molecular population inversion for various thickness of the gain layer. In contrast to conventional particle based spasers, we found negligible effect of the mutual dipole-dipole interaction on the spasing threshold and emission frequency.

\section{Acknowledgments}
This research was performed while the first author held a NRC Research Associateship Programs (USA) Award at the Air Force Research Laboratory. This work was also supported by AFRL Materials and Manufacturing Directorate Applied Metamaterials Program.

\appendix
\section{Appendix: Derivation of HMM Green function for $zz$ projection }

The simple expression for the Green's tensor components of dipole located in an infinite homogeneous background can be obtained analytically \cite{morse, tai}. However, situation becomes much more complex for dipoles located on top of stratified and uniaxial media, where tensor components need to account for reflection that occur at the surface of the HMM. The relevant reflected spatial domain Green's functions are usually formulated as Sommerfeld integrals \cite{maradudin} and accurate numerical evaluation of these integrals in the complex plane is critical. While this can be difficult in the most general of situations, in the context of our problem, sufficiently accurate results can be obtained by utilizing a short-distance asymptotic expansion of the Sommerfeld integrals valid for electrostatic approximation \cite{markel}. Assuming only nonradiative energy transfer between gain and HMM we can restrict ourself only with zero-order terms and neglect any possible radiation losses due to its extremely small contribution. The uniaxial nature of substrate also required corresponding modifications of derivations. Since the maximum emission strength can be obtained for the perpendicular orientation of molecular dipoles to the surface of HMM, we restrict our derivation only with $zz$ component
\begin{equation}
G_{zz} (R,Z)=-\frac{1}{4\pi k^2}\int_0^{\infty}\frac{q^3 J_0(qR)dq}{\chi_1(q)}R_{12}(q) exp[-\chi_1(q) Z],
\end{equation}
where $\chi_1(q)=\sqrt{q^2-k_1^2}$ with $k_1^2=k_0^2\epsilon_0$, $\epsilon_0$ - is a dielectric constant in top half-space and $k_0$ - wave vector in vacuum.  The reflection coefficient is defined as
\begin{equation}
R_{12}= \frac{\epsilon_x \chi_1(q)- \epsilon_0 \chi_2(q)}{\epsilon_x \chi_1(q)+ \epsilon_0 \chi_2(q)},
\end{equation}
where $\chi_2(q)=\sqrt{(\epsilon_x/\epsilon_z)q^2-k_0^2\epsilon_x}$ is effective parameter obtained for $p$-polarized wave in complex HMM \cite{sipe, sipe2}. The above integral diverges in upper limit. Following the way of derivations presented in \cite{markel}, we break it into two integrals: one from $0$ to $k_1$ and other from  $k_1$ to infinity. While the first integral becomes solvable, the divergence difficulties would remain in the second integral and can be treated by introducing a counter term:
\begin{eqnarray}
G_{zz} (R,Z)=-\frac{1}{4\pi k^2}\int_0^{k_1} f(q) J_0 (qR) dq exp(-\chi_1(q) Z) -
\nonumber\\
\frac{1}{4\pi k^2}\int_{k_1}^{\infty}[f(q)-g(q)]J_0(qR)exp(-\chi_1(q) Z)dq-
\nonumber\\
\frac{1}{4\pi k^2}\int_{k_1}^{\infty} g(q) J_0(qR) exp(-\chi_1(q) Z)dq,
\end{eqnarray}
where in above equation
\begin{equation}
f(q)=\frac{q^3}{\chi_1(q)}R_{12}(q)=\frac{q^3}{\chi_1(q)} \frac{\epsilon_x \chi_1(q)- \epsilon_0 \chi_2(q)}{\epsilon_x \chi_1(q)+ \epsilon_0 \chi_2(q)},
\end{equation}
and choice of counter term $g(q)$ is in principle arbitrary with single requirement to satisfy the following condition (while $q \Rightarrow \infty$ the contribution $f(q)-g(q)=O(1/q^{\alpha})$, where $\alpha>1$, is small enough). Further calculation reduces to prove that contributions from the first and second integral are sufficiently small and can be neglected while  the main contribution of the integral is contained in the last term which can be analytically taken. The form of function $g(q)$, which satisfies above mentioned conditions, we found in the form of an expansion of $R_{12}$ in powers of $1/q$ at large $q$
\begin{eqnarray}
g(q)\approx\frac{q^3}{\chi_1(q)}R_{12}(q \Rightarrow \infty),
\end{eqnarray}
where
\begin{eqnarray}
\label{R12}
R_{12}\approx\frac{\epsilon_x-\epsilon_0 \sqrt{\frac{\epsilon_x}{\epsilon_z}}}{\epsilon_x+\epsilon_0 \sqrt{\frac{\epsilon_x}{\epsilon_z}}} +
\nonumber\\
\frac{k_1^2\epsilon_x \sqrt{\frac{\epsilon_x}{\epsilon_z}}(\epsilon_z-\epsilon_0)}{q^2 (\epsilon_x + \epsilon_0
\sqrt{\frac{\epsilon_x}{\epsilon_z}})^2} + O(1/q^n)...,
\end{eqnarray}
Equation \ref{R12} can be simplified by introducing a new parameter
\begin{eqnarray}
S= \frac{\epsilon_x-\epsilon_0 \sqrt{\frac{\epsilon_x}{\epsilon_z}}}{\epsilon_x+\epsilon_0 \sqrt{\frac{\epsilon_x}{\epsilon_z}}}=
\frac{\epsilon-\tilde\epsilon}{\epsilon+\tilde\epsilon},
\end{eqnarray}
where $\epsilon=\epsilon_x/\epsilon_0$  and $\tilde\epsilon=\sqrt{\frac{\epsilon_x}{\epsilon_z}}$.
\begin{eqnarray}
g(q)\approx\frac{q^3}{\chi_1(q)}\left[ S + \frac{k_1^2 \epsilon_x (1-S)^2 (\epsilon_z-\epsilon_0)}{4 q^2 \tilde \epsilon}\right],
\end{eqnarray}
That integral can be evaluated analytically \cite{markel} and, finally, we can represent the Green's function as an short-distance expansion in powers of $(k_1 L)$ retaining only on zero order.
\begin{eqnarray}
G= \frac{1}{4\pi k^2 L^3} \sum_{l=0}^{\infty} (k_1 L)^l K^{(l)},
\end{eqnarray}
where zero order of expansion
\begin{eqnarray}
 K^{(0)}= S \left[3\frac{Z^2}{L^2} -1\right],
 \end{eqnarray}
would corresponds to the non-radiative (non-emissive) part of the Green's function
\begin{eqnarray}
G_{zz}^{nr} = \frac{S}{4 \pi k^2 L^3} \left[3\frac{Z^2}{L^2} -1\right],
 \end{eqnarray}
 Further calculations of higher orders of the Green's function expansion are considerably more complicated and in the context of our problem generally redundant since for HMMs all radiative losses are negligible. Those calculations have been carried out but are not presented here.

\end{document}